\documentclass[aps,preprint,showpacs]{revtex4}

\usepackage{graphicx}

\usepackage{hyperref}  

\usepackage{amsfonts}
\usepackage{amsbsy}
\usepackage{amssymb}

\begin{document}

\title{The standard map: \\ 
From Boltzmann-Gibbs statistics to Tsallis statistics}

% article:
\author{Ugur Tirnakli$^{1,}$}
 \email{ugur.tirnakli@ege.edu.tr}
\author{Ernesto P. Borges$^{2,}$}
 \email{ernesto@ufba.br}

\affiliation{
$^1$Department of Physics, Faculty of Science, Ege University, 35100 Izmir, Turkey\\
$^2$Instituto de F\'isica, Universidade Federal da Bahia, Rua Bar\~ao de Jeremoabo, 
40170-115 Salvador-BA, Brasil \\}

%\date{\today}

%%%%%%%%%%%%%%%%%%%%%
\begin{abstract}
%%%%%%%%%%%%%%%%%%%%%
As well known, Boltzmann-Gibbs statistics is the correct way of
thermostatistically approaching ergodic systems. On the other hand,
nontrivial ergodicity breakdown and strong correlations typically
drag the system into out-of-equilibrium states where Boltzmann-Gibbs
statistics fails. For a wide class of such systems, it has been shown
in recent years that the correct approach is to use Tsallis statistics
instead. Here we show how the dynamics of the paradigmatic conservative
(area-preserving) standard map exhibits, in an exceptionally clear
manner, the crossing from one statistics to the other. Our results unambiguously
illustrate the domains of validity of both Boltzmann-Gibbs and Tsallis
statistics.
\end{abstract}


\maketitle

%%%%%%%%%%%%%%%%%%%%%%%%%%%%%%%%%%%
\section{Introduction}
%%%%%%%%%%%%%%%%%%%%%%%%%%%%%%%%%%%

Exponential and Gaussian distributions are signatures of 
the Boltzmann-Gibbs statistical mechanics. These distributions are
those that maximise the Boltzmann-Gibbs entropy and ensure the equilibrium state.
The Maxwell distribution is an instance of the equilibrium 
distribution for the velocities of molecules in an ideal gas.
The underlying mathematical reason for this is the existence of the 
standard Central Limit Theorem (CLT) \cite{CLT}. 
On the other hand, due to ergodicity breaking, some systems remain indefinitely trapped 
into non-exponential and non-Gaussian distributions, and thus achieve 
out-of-equilibrium quasi-stationary states. 
The $q$-exponential and the $q$-Gaussian distributions are 
functions associated with some of these quasi-stationary states 
and they are the maximising distributions for the non-additive Tsallis entropy 
given by $S_{q} \equiv k \left(1- \sum_i p_i^q\right)/ \left(q-1\right)$ \cite{tsallis88}. 
This feature permits to describe these special non-equilibrium states
with the same formal framework of the equilibrium thermostatistics, known as 
Tsallis statistics \cite{tsallisbook}, 
and this general picture is reduced to the equilibrium one 
if the parameter $q$ attains a special limiting value ($q\rightarrow 1$). 
In this case, the underlying mathematical mechanism is the generalized CLT \cite{qCLT1,qCLT2}, 
which states that the stable limit distributions for a certain class of systems in such quasi-stationary 
states are $q$-Gaussians. 
Therefore, the role of $q$-Gaussians in Tsallis statistics is basically the same as 
that of Gaussians in Boltzmann-Gibbs statistics. 
In this work we show, for the first time, that these two cases coexist 
in the classical standard map, and discuss the necessary conditions under 
which one case prevails over the other one. This neatly illustrates the
respective domains of validity of Boltzmann-Gibbs and of Tsallis
statistics.

Non-Gaussian distributions, particularly $q$-Gaussians, have been observed in nature in several experimental, 
observational and model systems \cite{tsallisbook}. Impressive experimental examples include 
(i)~a high dimensional dissipative system where the probability densities of velocity differences measured in 
a Couette-Taylor experiment for a fully developed turbulence 
regime \cite{beck-lewis-swinney-2001,tsallis-borges-baldovin-2002}, 
(ii)~transport properties of cold atoms in dissipative optical lattices \cite{renzoni1,renzoni2} and 
(iii)~transverse momentum spectra of hadrons at LHC experiments \cite{wilk}. 
As observational works, for small bodies in the Solar System, particularly asteroid rotation periods and 
diameters \cite{betzler-borges-2012} and distribution of meteor showers \cite{betzler-borges-2014} can be given. 
At a larger scale, the rotation curve for the M33 Triangulum Galaxy has been successfully analyzed in the same 
sense \cite{cardone-leubner-delpopolo-2011}. 
Among model systems, one of the paradigmatic dissipative low dimensional model, the logistic map, has been 
numerically investigated and $q$-Gaussians have been found as the chaos threshold is approached using the 
band splitting structure obeying the Huberman-Rudnick scaling law \cite{tibet2005,tibet2009,afsar1}.

$q$-Gaussians have also been recently observed in a conservative high dimensional model  \cite{cirto-assis-tsallis-2014}.
In the $\alpha$-XY model, i.e., a system of $N$ classical localized planar rotators with two-body interactions and periodic 
boundary conditions, the potential is assumed to decay with distance as $1/r^\alpha$, and $\alpha \ge 0$ is the 
parameter that controls the range of the interactions, short-range for $\alpha/d > 1$, and long-range for 
$0 \le \alpha/d \le 1$ ($d$ is the spatial dimensionality of the system).

Recently a generalization of the conservative one-dimensional Fermi-Pasta-Ulam model, properly modified to account for 
linear and nonlinear long-range interactions, has been analyzed. The range of the interactions is controlled in the 
same way as for the $\alpha$-XY model just mentioned. 
Ordinary Gaussians are observed when short-range interactions ($\alpha>1$) are present, and $q$-Gaussians are observed 
when long-range interactions ($0\le \alpha \le 1$) are present \cite{Christodoulidi-Tsallis-Bountis-2014}. 
It has been found that the maximal Lyapunov exponent $\lambda$ asymptotically decreases as $N^{-\kappa(\alpha)}$, 
in a rather similar way of that observed in \cite{anteneodo-tsallis-1998} for the $\alpha$-XY model and the $q$-Gaussian 
distributions that emerge are characterized by the parameter $q$ that depends on $\alpha$.

All these systems appear to share in common the following scenario:
ergodicity in a region is characterized by the largest Lyapunov exponent $\lambda$ and two regimes shall be distinguished 
in the thermodynamic limit (number of particles $N\to\infty$).
(i) Strongly chaotic regime corresponds to a large positive Lyapunov exponent, where the system is ergodic. 
The dynamics of the system evolves to an equilibrium state described by Boltzmann-Gibbs statistical mechanics, 
with exponential or Gaussian distributions (according to the considered dynamical variable);
(ii) Weakly chaotic regime corresponds to a very small positive Lyapunov exponent ($\lambda \approx 0$), 
where the system behaves for a very long time as non-ergodic. Distributions of the dynamical variables are not 
exponential or Gaussians, and the Boltzmann-Gibbs framework is not suitable for this case. 
$q$-Gaussian distributions have been observed for this case by proper time and ensemble averages \cite{ruiz}. 
These distributions are obtained by maximisation of the nonadditive entropy $S_q$ \cite{tsallis88,ct-levy,ct-prato}, 
which is a strong indication that these systems are connected to nonextensive statistical mechanics \cite{tsallisbook}. 
They may be written as $P(u)\propto \exp_q(-B u^2)$, where $B>0$ is the Lagrange parameter and 
the $q$-exponential is given by $\exp_q u = [1+(1-q) u]^{1/(1-q)}_+$, with $[A]_+ \equiv \max\{0,A\}$ and its 
inverse, $q$-logarithm, is defined by $\ln_q u = (u^{1-q}-1)/(1-q)$.
The ordinary Gaussian, exponential and logarithm functions are
respectively recovered in the limit $q \to 1$.

In this paper we consider the standard map, that is a paradigmatic low dimensional conservative (area-preserving) model, 
and we follow the averaging procedure originally used for the logistic map, as described 
in \cite{tibet2005,tibet2009,afsar1}. 
As will be discussed in detail below, this paradigmatic model offers an excellent medium for us to analyse 
both regimes explained above and to establish a connection between these regimes where the system is 
ergodic and non-ergodic.

%%%%%%%%%%%%%%%%%%%%%%%%%%%%%%%%%%%%%%%%%%%%%%%%%%%%%%%
\section{\label{sec:q-gaussians} The Model and Results}
%%%%%%%%%%%%%%%%%%%%%%%%%%%%%%%%%%%%%%%%%%%%%%%%%%%%%%%

\begin{figure}[ht]
\centering
 \includegraphics[scale=0.53]{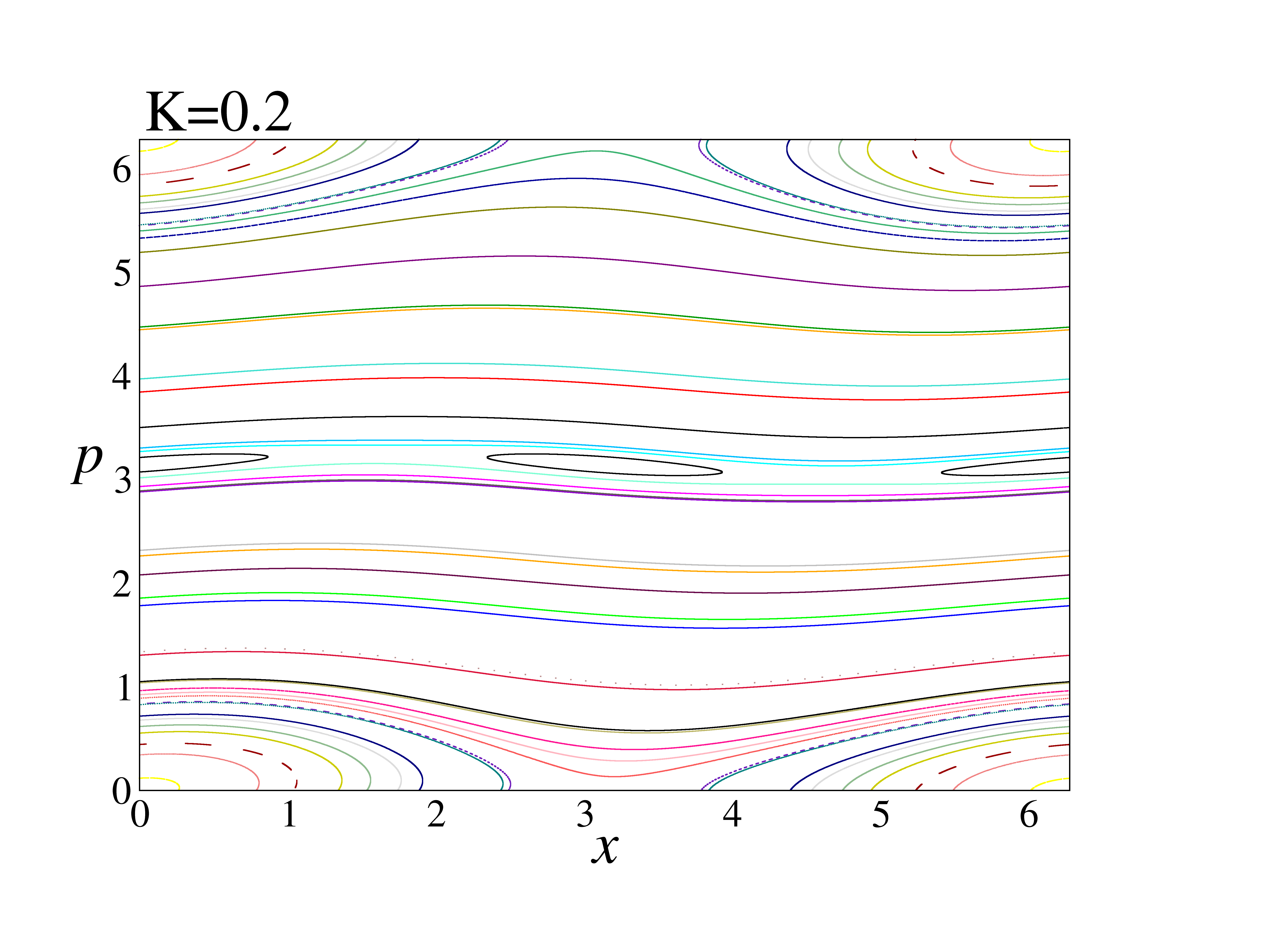}
 \includegraphics[scale=0.53]{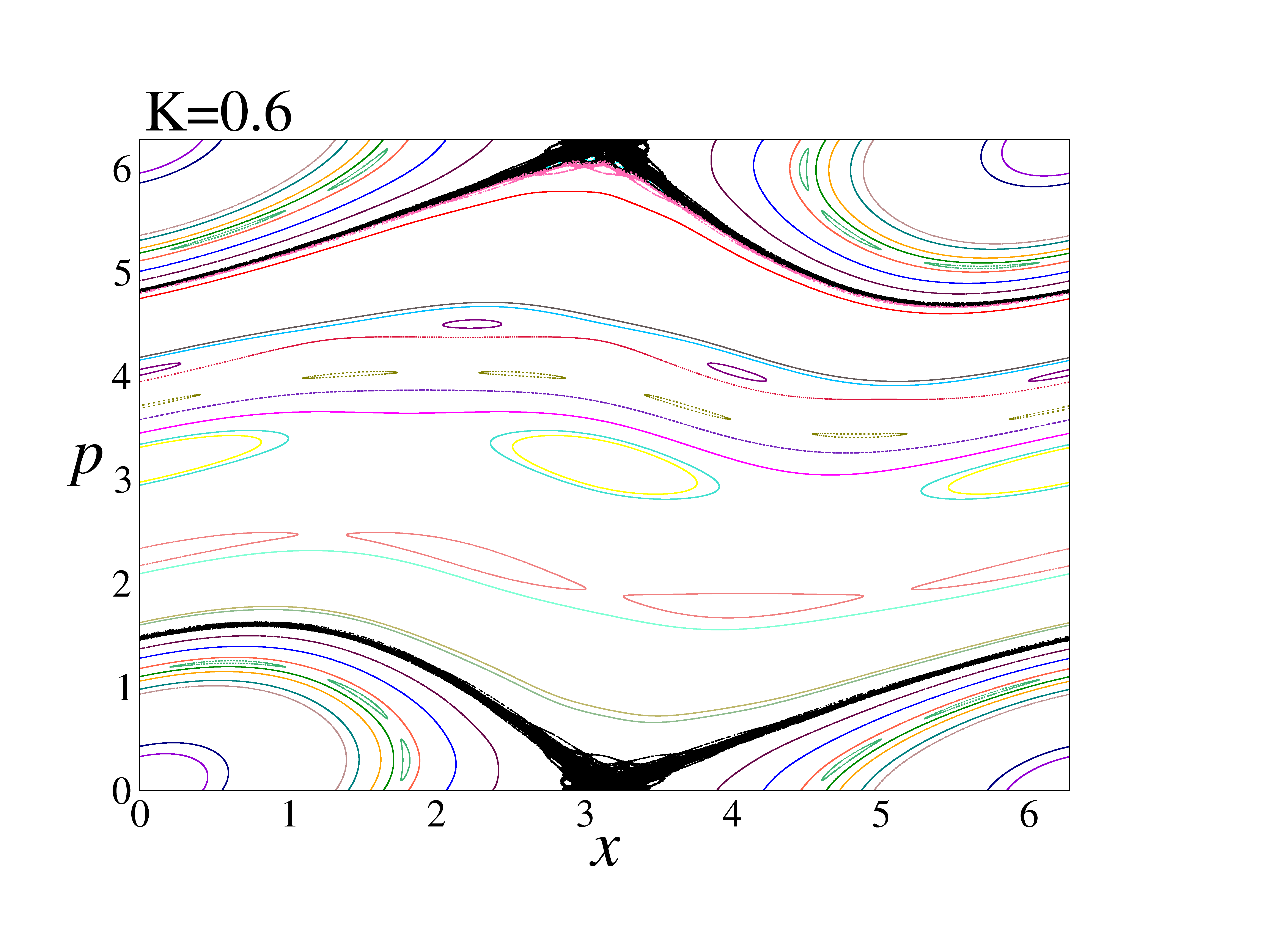}
 \includegraphics[scale=0.53]{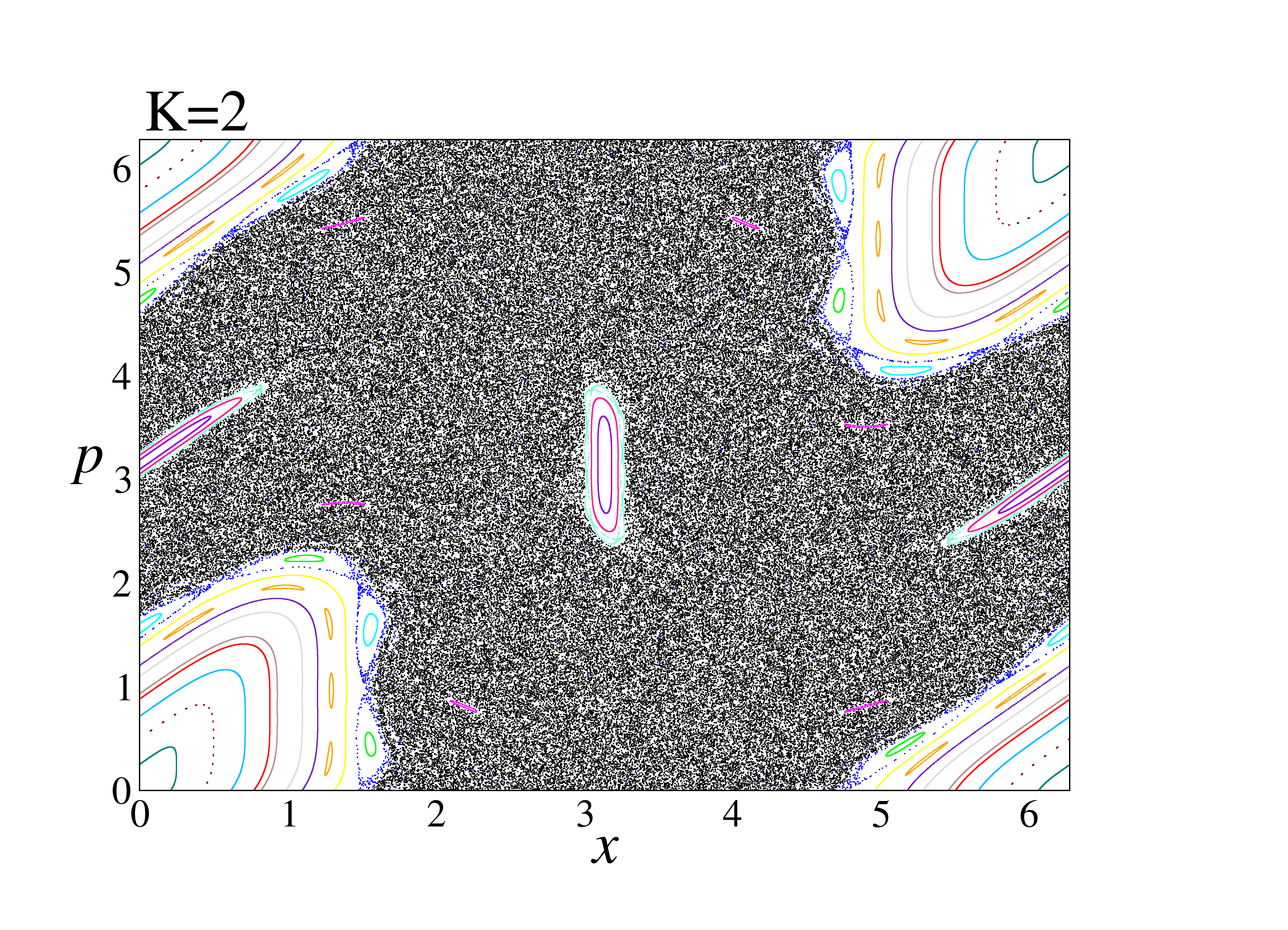}
 \includegraphics[scale=0.53]{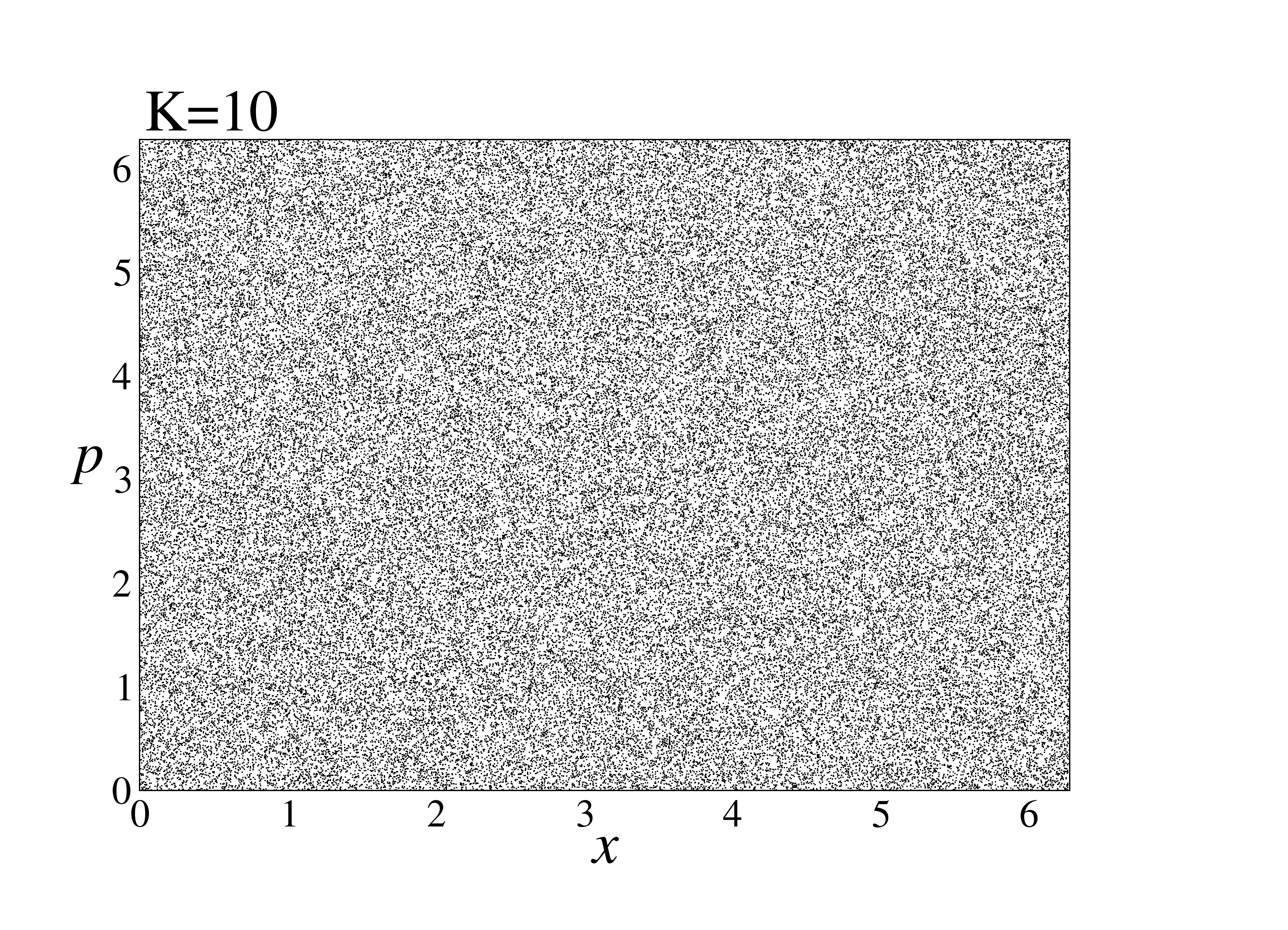}
 \caption{\label{fig:phasediagram}
(Color online) Phase portrait of the standard map for 4 representative $K$ values. 
In each case, black dots represents the region of chaotic sea in the available phase space 
and all other colors represent different stability islands.}
\end{figure}

The standard map is defined as \cite{chirikov,zaslavsky1}
\begin{equation}
 \label{eq:stdmap}
 p_{i+1} = p_i - K \sin x_i\qquad;\qquad x_{i+1}=x_i+p_{i+1}
\end{equation}
where $p$ and $x$ are taken as modulo $2\pi$. 
This map has very rich properties depending on the map parameter $K$. 
Here, we will focus on four representative cases whose phase portraits are given 
in Fig.~\ref{fig:phasediagram}. 
The two extreme cases are $K=0.2$ and $K=10$, one of which represents the domination 
of the phase space with the stability islands and the other is clearly an example of the 
invasion of the full phase space by the chaotic sea. On the other hand, the other two 
cases in between, namely, $K=0.6$ and $K=2$, are good examples in order to see how 
these regions with stability islands and chaotic sea merge in the available phase space.  
It is clear that if the system starts from an initial condition located on one of the archipelagos 
(given by the same color), it will stay forever in the same archipelago, whereas if it starts 
from somewhere in the chaotic sea, the iterates will cover the whole chaotic region.

%%%%%%%%%%%%%%%%%%%%%%%%%%%%%%%
%%%%%%%%%%%%%%%%%%%%%%%%%%%%%%%%%%%%%

\begin{figure}[ht]
\centering
\includegraphics[scale=0.7]{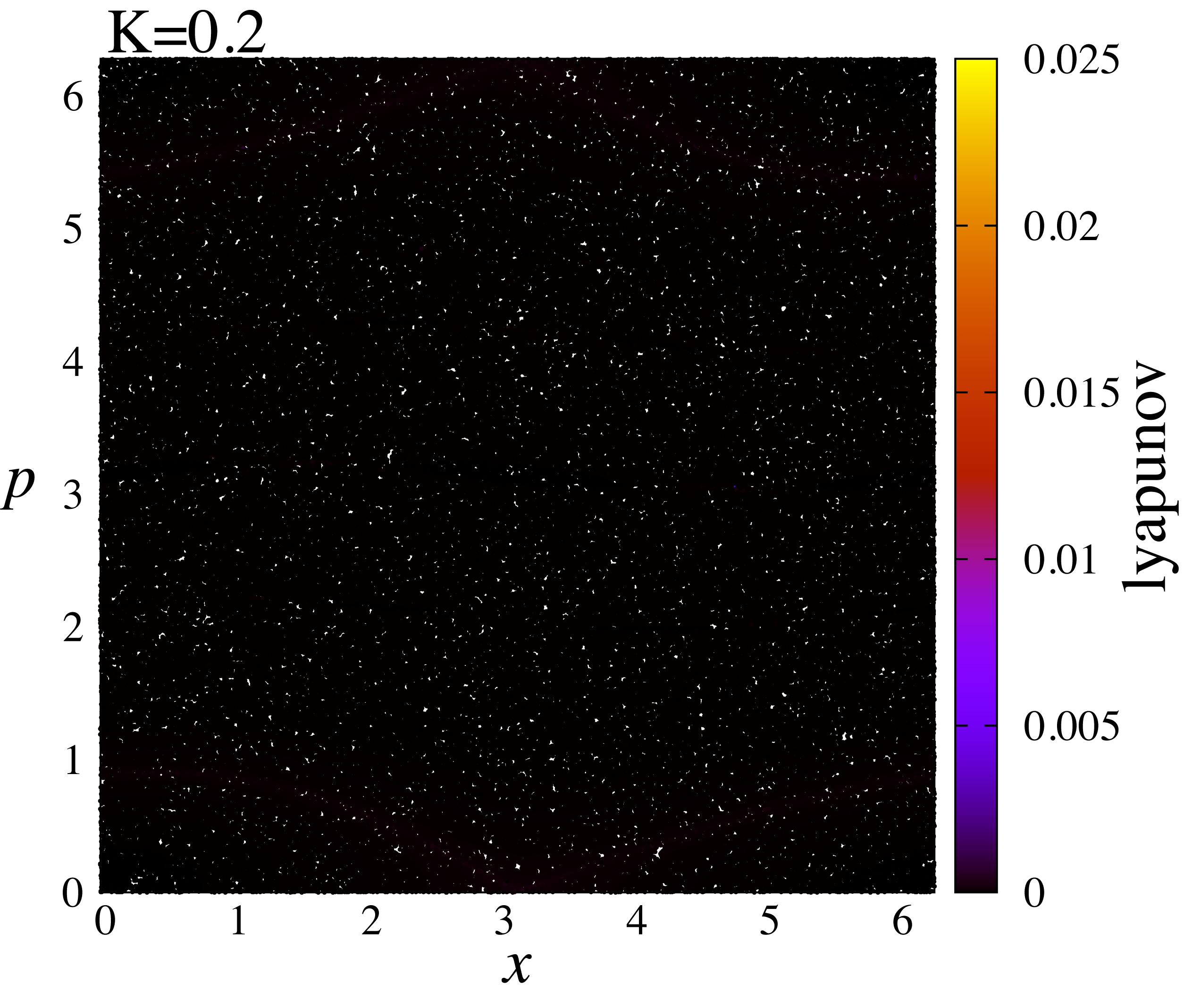}
\includegraphics[scale=0.7]{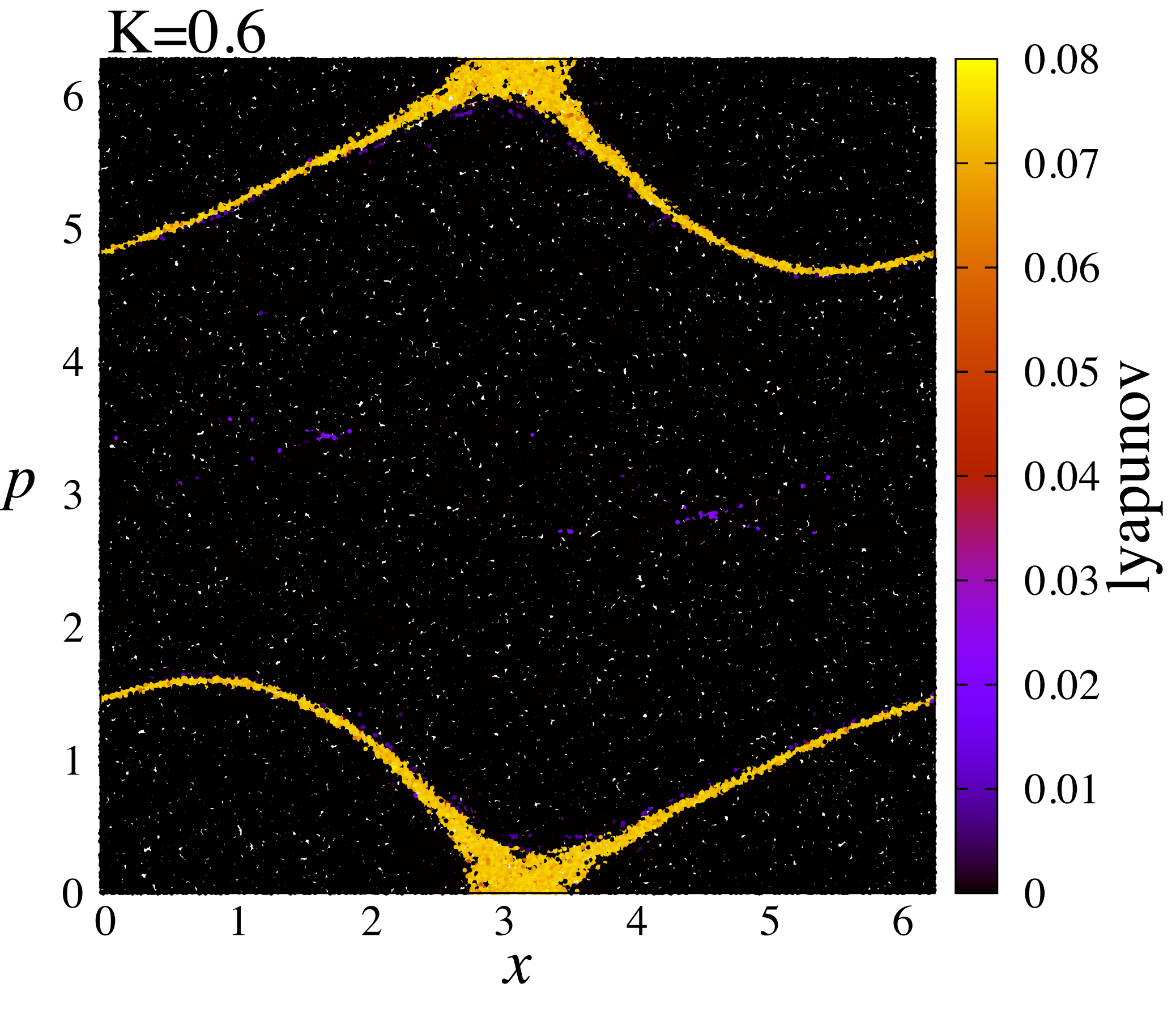}
\includegraphics[scale=0.7]{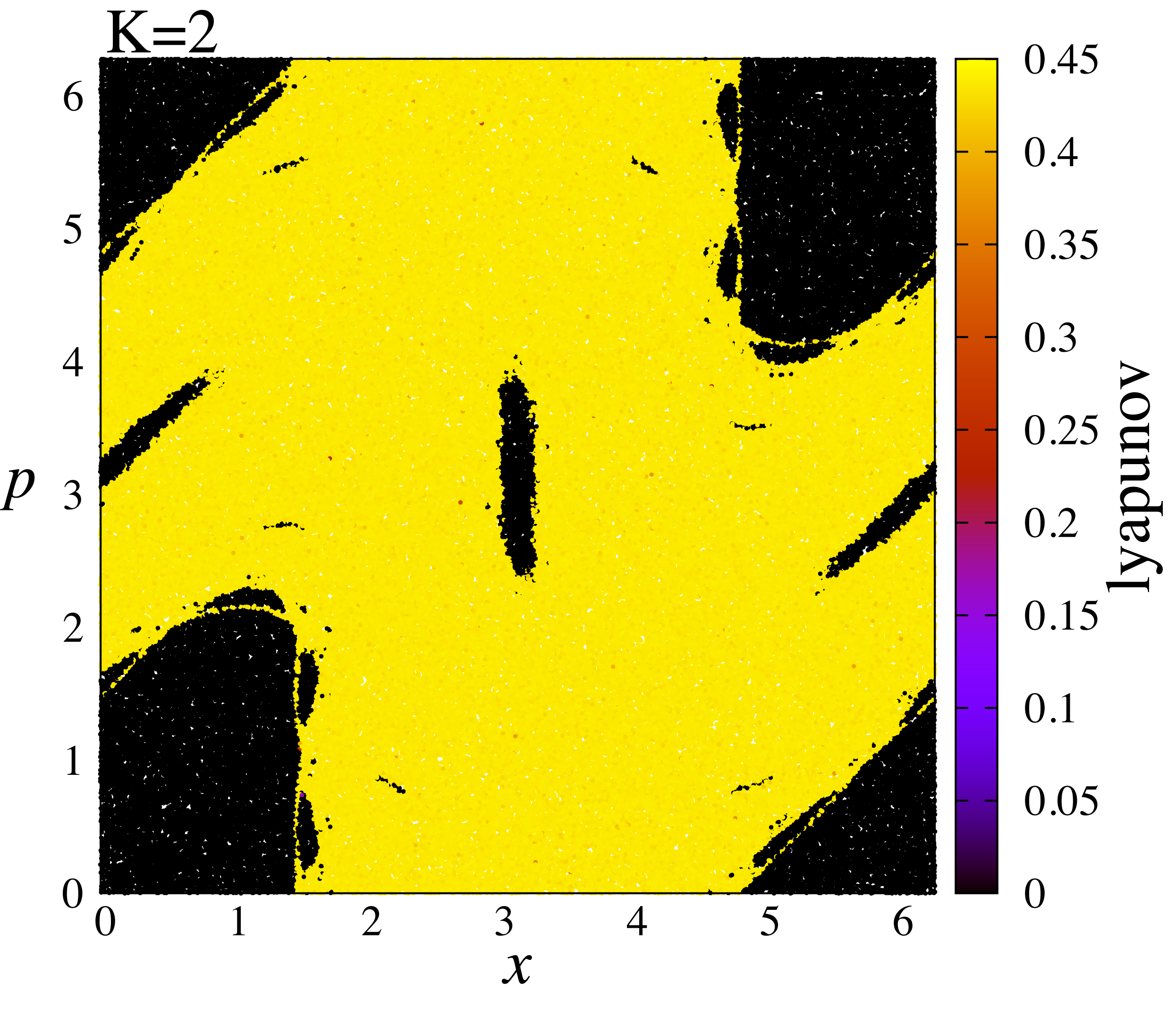}
\includegraphics[scale=0.7]{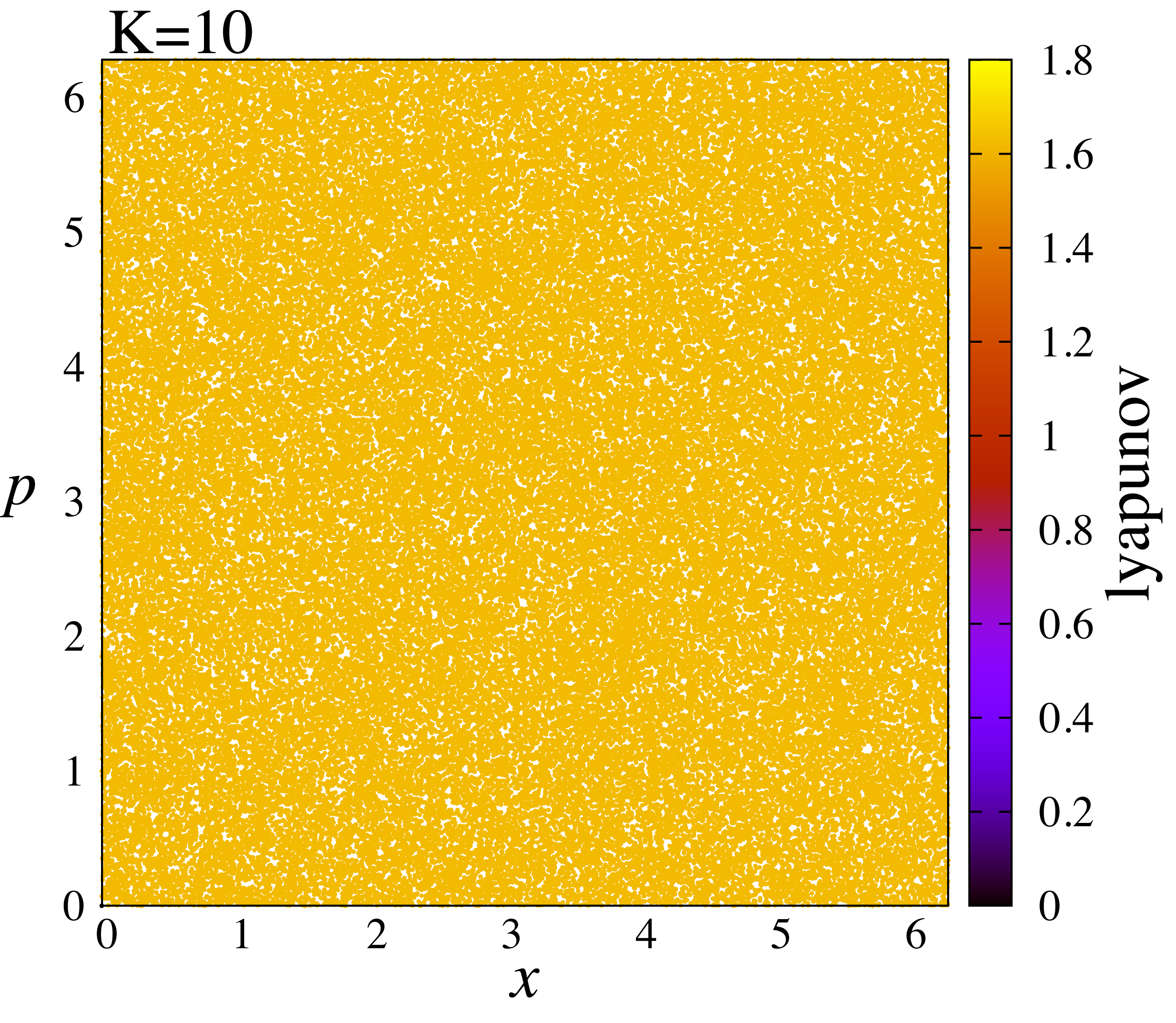}
\caption{\label{fig:lyaphasediagram}
(Color online) Lyapunov exponent results of the phase portrait of the standard map. 
The same representative $K$ values are used. 
For each case, Lyapunov exponents are calculated for 200000 initial conditions. 
In the calculation, each initial condition is iterated $10^7$ times.}
\end{figure}

At this point, we need to calculate the largest Lyapunov exponent of these cases using the Benettin 
algorithm \cite{benettin} but this calculation is to be done very carefully. Generally, calculating the 
Lyapunov exponent by taking an ensemble average would not be exactly correct here since the 
contributions coming from the initial conditions of stability islands are much smaller than the ones 
coming from the chaotic sea. Therefore, making an ensemble average would not reflect the correct 
behaviour of the system. In order to reflect the correct behaviour, we prefer to plot the largest 
Lyapunov exponents as given in Fig.~\ref{fig:lyaphasediagram}, where we calculate the exponent of 
each initial condition separately over the whole phase space and the magnitude of the exponents 
are given by a color map. 
As seen in the figure, the case $K = 0.2$ represents a Lyapunov spectrum in which all results are 
extremely close to zero (black dots), whereas the case $K=10$ conversely exhibits a spectrum 
where all results are largely positive (yellowish dots). This means that, in the former case (latter case), 
the whole phase space is dominated by the stability islands (chaotic sea). 
On the other hand, the other two cases, $K=0.6$ and $K=2$, are good examples where the phase 
space consists of both stability islands and chaotic sea. 
This way of representing the Lyapunov spectrum allows us to see clearly the portions of the whole 
phase space where the system is ergodic and non-ergodic for a given $K$ value.

%%%%%%%%%%%%%%%%%%%%%%%%%%%%%%%
%%%%%%%%%%%%%%%%%%%%%%%%%%%%%%%%%%%%%

Now we can analyze the limit distributions of the standard map for these representative $K$ values. 
We define the variable
\begin{equation}
 \label{eq:y}
 y:=\sum_{i=1}^T (x_i - \langle x \rangle ),
\end{equation}
where the average $\langle \cdots \rangle$ is calculated as time average 
taken over not only a large number of $T$ iterations, but also a large number of 
$M$ randomly chosen initial values, namely,
\begin{equation}
 \label{eq:average}
 \langle x \rangle = \frac{1}{M}\frac{1}{T}\sum_{j=1}^M \sum_{i=1}^T x_i^{(j)}, 
\end{equation}
and calculate the probability distribution of $y$, namely $P(y)$, for any given $K$ parameter.

\begin{figure}[ht]
 \begin{center}
 \includegraphics[width=0.9\columnwidth,keepaspectratio,clip=]{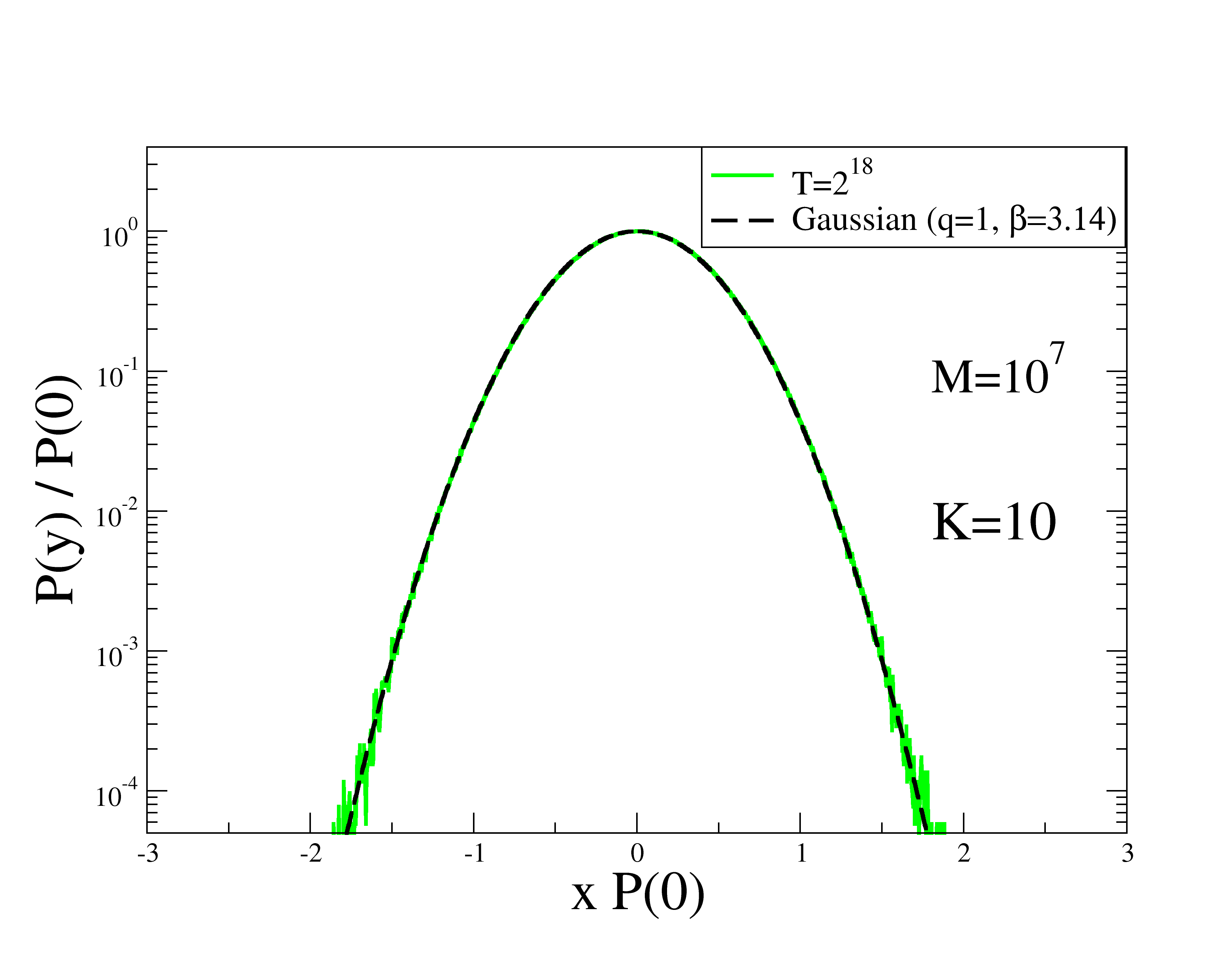}
%  \fbox{Missing figure}
 \end{center}
 \caption{\label{fig:distrib.k.10}
(Color online) Normalized probability distribution function for the case $K=10$ with $T=2^{18}$.}
\end{figure}

Let us start with the result of the case $K = 10$, where the probability distribution is expected 
to be a Gaussian since for this case the phase space is totally a chaotic sea, which makes the 
whole system ergodic. The result is given in Fig.~\ref{fig:distrib.k.10} where a clear Gaussian 
is easily seen as expected. It should also be noted that the stable limit distribution is 
obtained quickly.

%%%%%%%%%%%%%%%%%%%%%%%%%%
%%%%%%%%%%%%%%%%%%%%%%%%%%%

\begin{figure}[ht]
 \begin{center}
 \includegraphics[width=0.9\columnwidth,keepaspectratio,clip=]{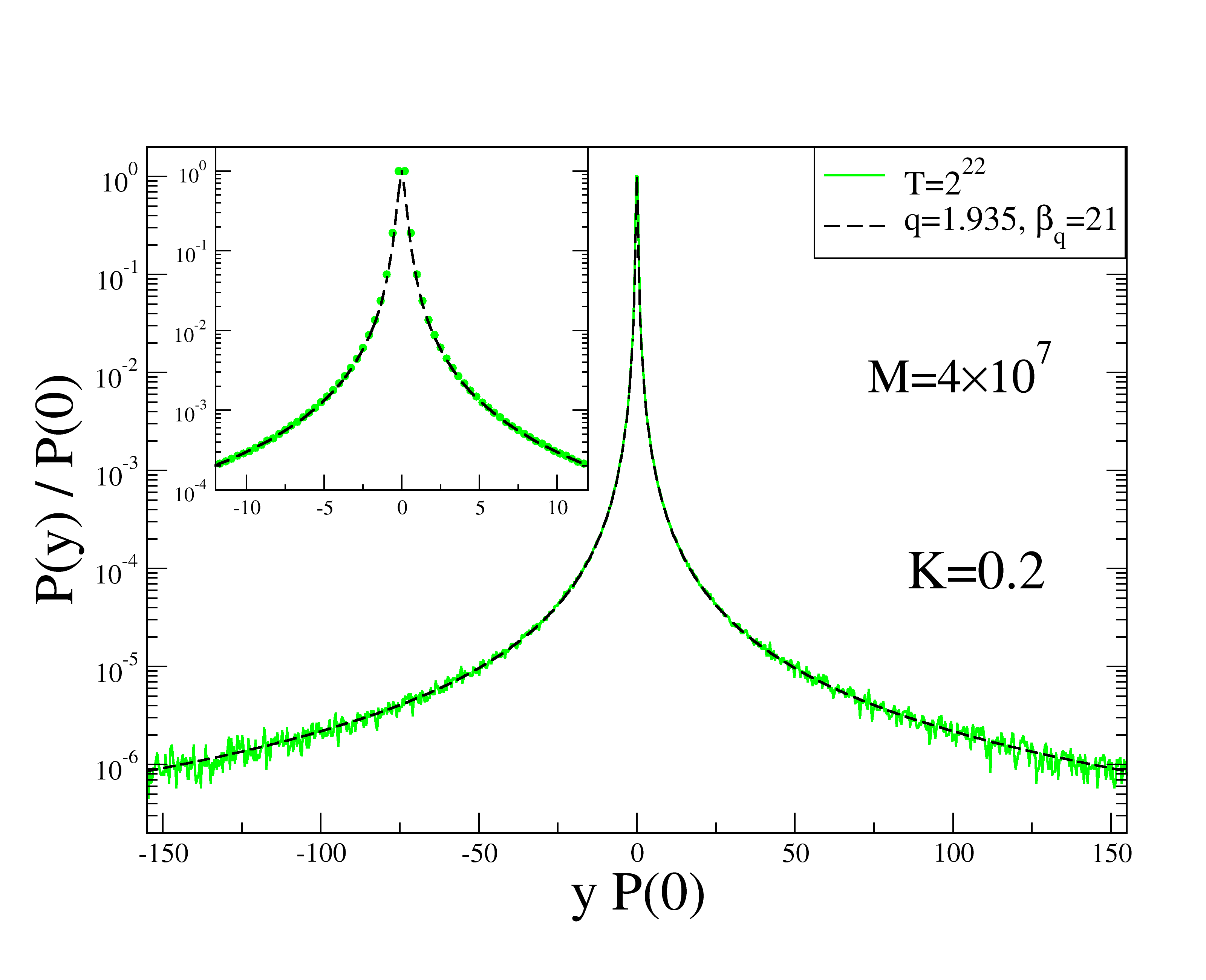}
 \end{center}
 \caption{\label{fig:distrib.k.02}
(Color online) Normalized probability distribution function for the case $K=0.2$ with $T=2^{22}$. 
In the Inset, the central part is zoomed for a better visualization.}
\end{figure}

\begin{figure}[ht]
 \begin{center}
 \includegraphics[width=0.9\columnwidth,keepaspectratio,clip=]{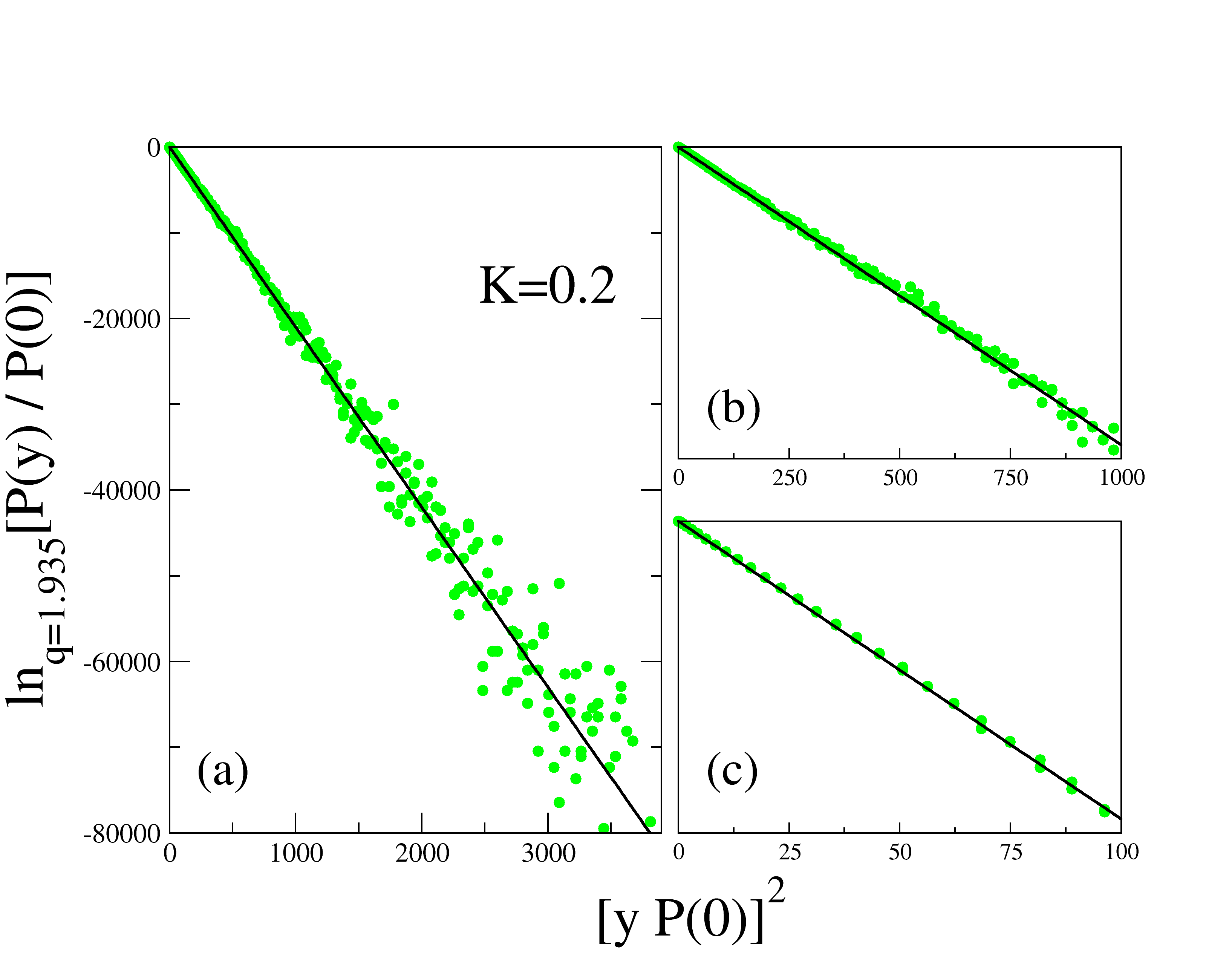}
 \end{center}
 \caption{\label{fig:lndistrib.k.02}
(Color online) $q$-logarithmic representation of the normalized probability distribution (a)~for the tails, 
(b)~for the intermediate region and (c)~for the central part of the case $K=0.2$. }
\end{figure}

Now we can investigate the case $K = 0.2$, where the probability distribution is expected 
to be a non-Gaussian due to the change of the phase space from being totally chaotic to 
totally consists of stability islands, which makes the whole system non-ergodic. 
The result is given in Fig.~\ref{fig:distrib.k.02} where, instead of a Gaussian, now a clear 
$q$-Gaussian is observed with $q=1.935$. In this case, the stable limit distribution happens 
to be achieved slowly but at the level of $T=2^{22}$ it has already been reached. 
We have checked it with $T=2^{23}$ and verified that the distribution does not change in the displayed region. 
We also plot the same data as $q$-logarithm of the probability distribution in Fig.~\ref{fig:lndistrib.k.02} 
in order to see whether it is a straight line or not. For three different regions (i.e., the region 
including the tails, intermediate region and the central part), the straight line is well 
approached.  
The fact that we observe straight lines in all scales excludes other
distributions that are asymptotic power laws, like L\'evy distributions.

%%%%%%%%%%%%%%%%%%%%%%%%%%%%%%
%%%%%%%%%%%%%%%%%%%%%%%%%%%%%%

%%%%%%%%%%%%%%%%%%%%%%%%%%%%%%
%%%%%%%%%%%%%%%%%%%%%%%%%%%%%%

\begin{figure}[ht]
\centering
 \begin{minipage}[h]{0.8\linewidth}
  \includegraphics[width=\linewidth]{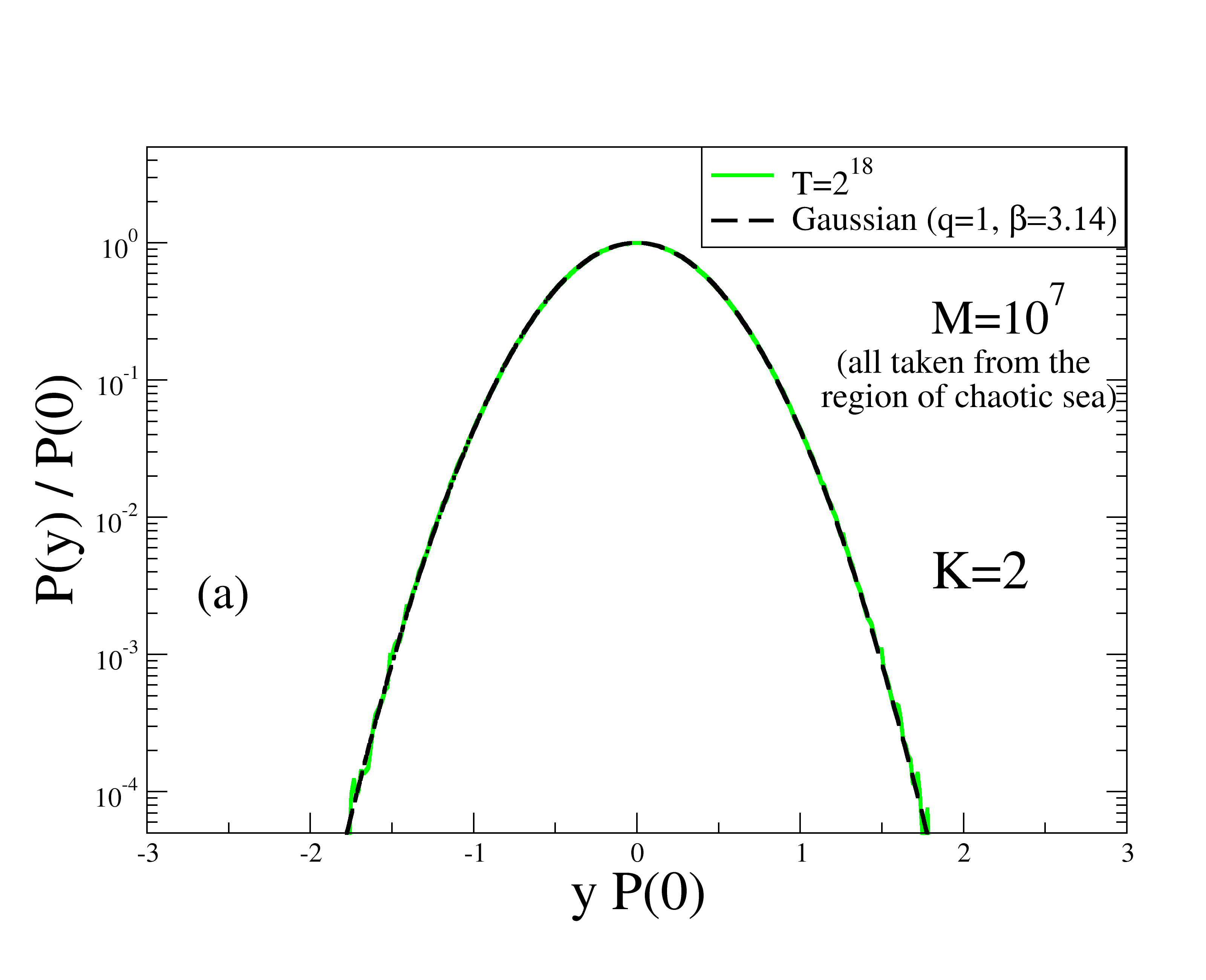}
 \end{minipage}
 \begin{minipage}[h]{0.8\linewidth}
  \includegraphics[width=\linewidth]{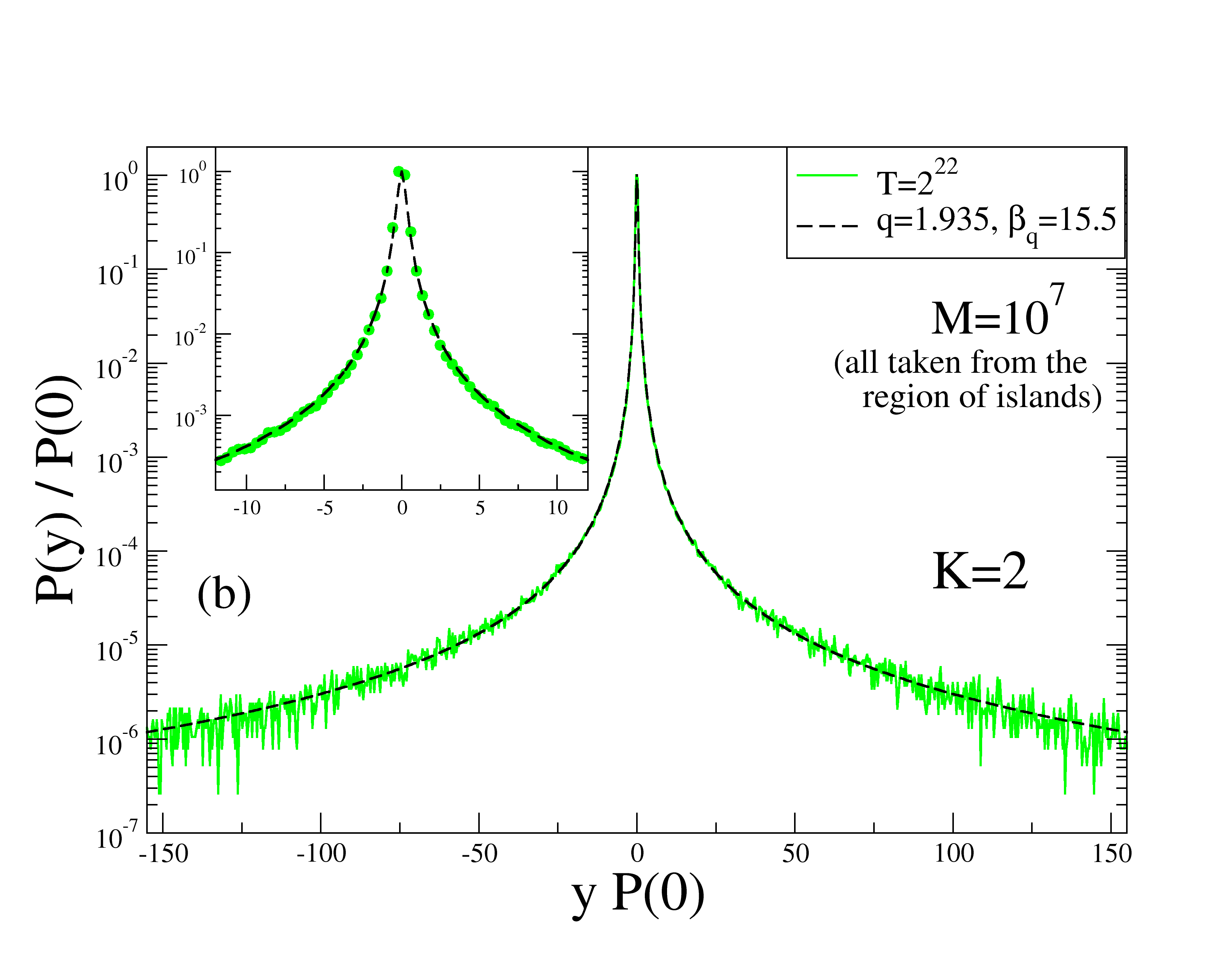}
 \end{minipage} 
 \caption{\label{fig:k2}
(Color online) Normalized probability distribution function for the case $K=2$. 
In the calculations, all initial conditions are taken from the region of (a)~chaotic sea 
and (b)~stability islands.}
\end{figure}

In order to better illustrate this tendency, we also perform another test by taking a $K$ value 
where the stability islands and chaotic sea coexist, i.e., the case $K=2$. 
For this case, it is evident from Fig.~\ref{fig:lyaphasediagram} that the region of chaotic sea 
with large positive Lyapunov exponents and the region of stability islands with Lyapunov 
exponents close to zero can easily be detected. This means that the system is ergodic within 
some portion of the phase space, whereas it is indeed non-ergodic within some other portion. 
Therefore we can check our previous findings using these portions separately. If we use initial conditions all taken 
from the portion where the system is ergodic (non-ergodic), we expect to see the same distribution function 
we have found before, namely the Gaussian ($q$-Gaussian with $q=1.935$). 
In fact, this is exactly what we see in  Fig.~\ref{fig:k2}, which nicely corroborates our results 
given in Fig.~\ref{fig:distrib.k.10} and Fig.~\ref{fig:distrib.k.02}.

%%%%%%%%%%%%%%%%%%%%%%%%%%%%%%%
%%%%%%%%%%%%%%%%%%%%%%%%%%%%%%%%%%%%%

\begin{figure}[ht]
\centering
 \begin{minipage}[h]{0.8\linewidth}
  \includegraphics[width=\linewidth]{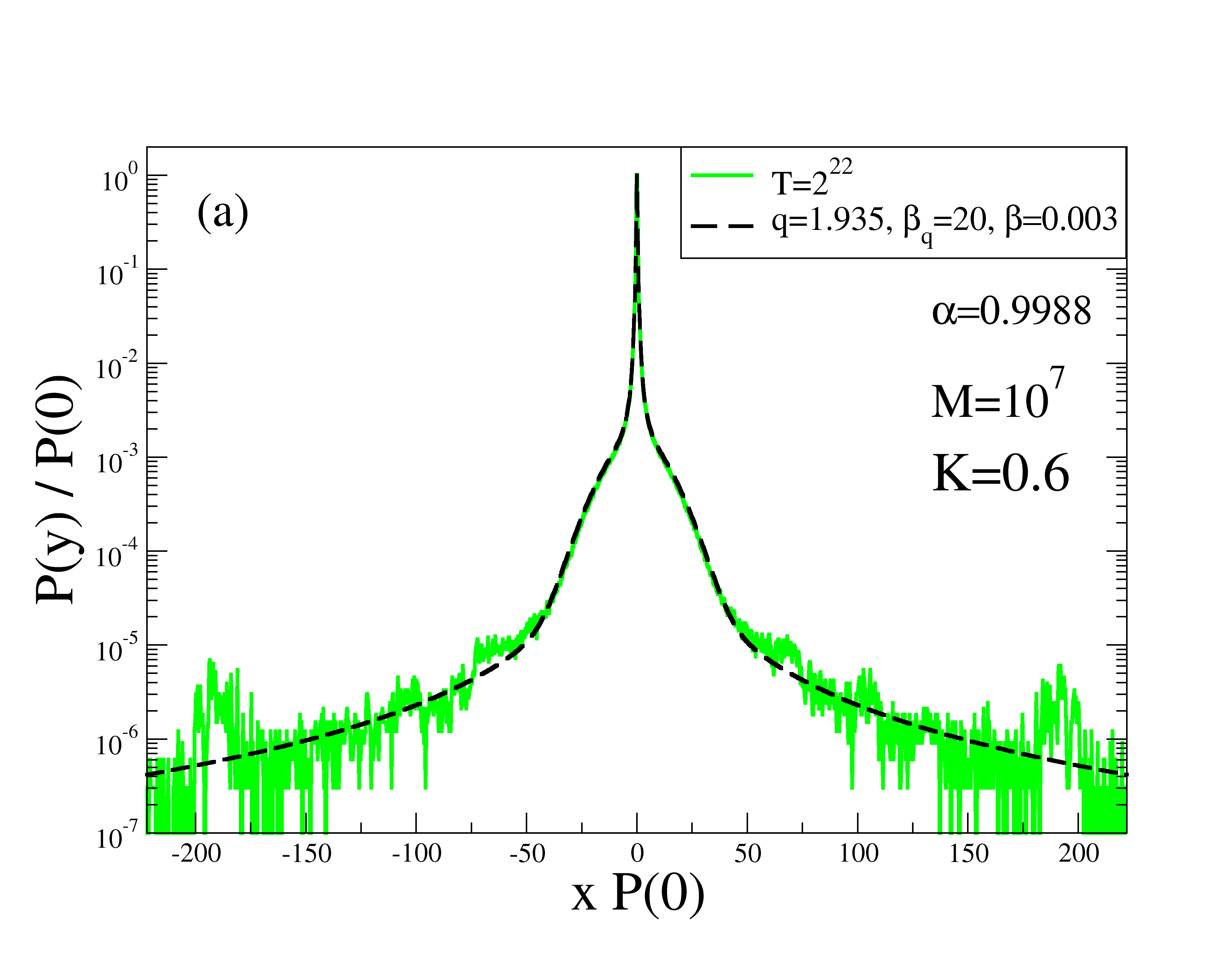}
 \end{minipage}
 \begin{minipage}[h]{0.8\linewidth}
  \includegraphics[width=\linewidth]{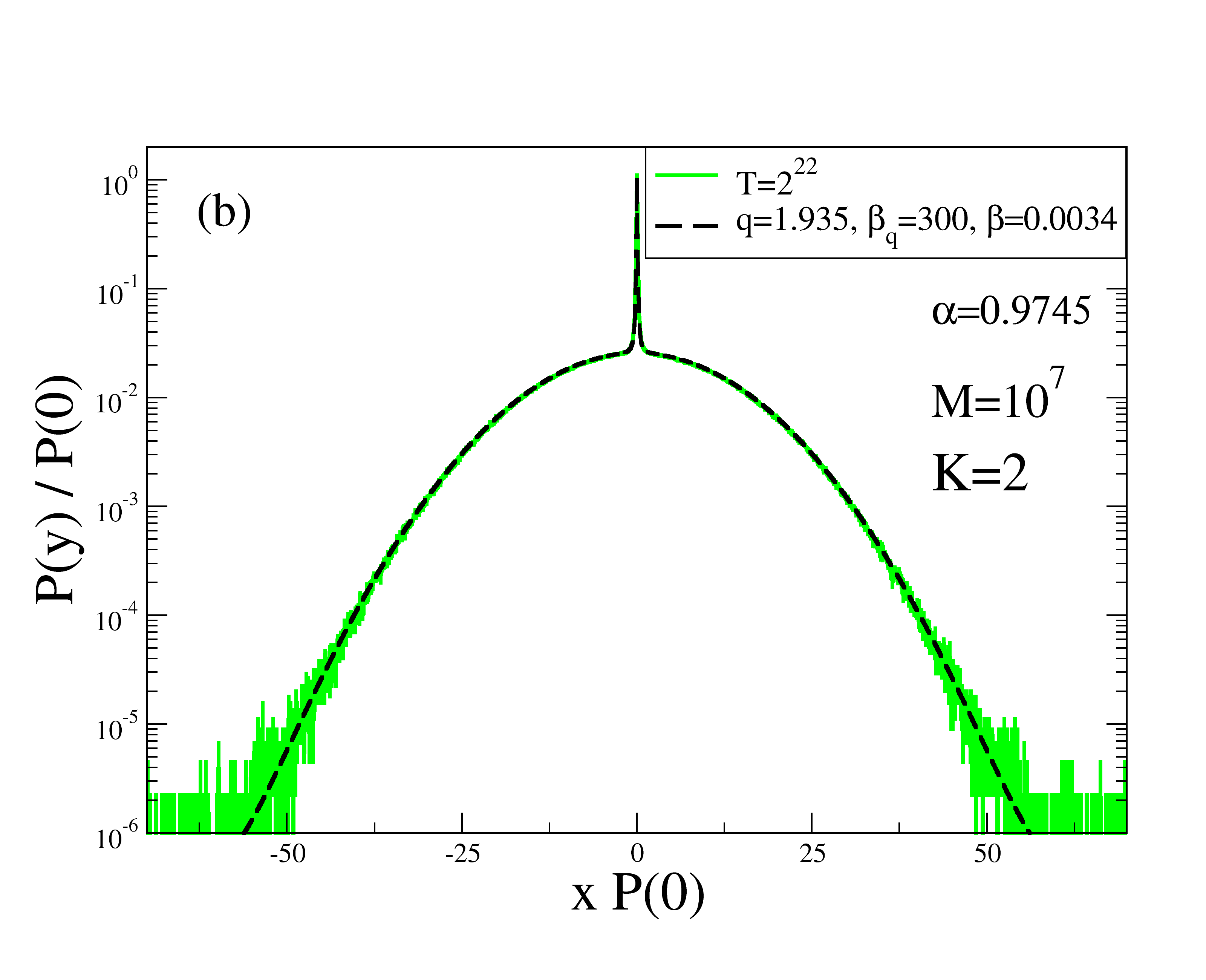}
 \end{minipage} 
\caption{\label{fig:k08-k2}
(Color online) Normalized probability distribution function for the cases (a)~$K=0.6$ and (b)~$K=2$. 
In the calculations, initial conditions are randomly taken from the whole available phase space. }
\end{figure}

Finally we will be interested in another interesting question: what happens to the probability 
distribution if we do not take the portions of the phase space separately where the system is ergodic and non-ergodic 
but consider initial conditions coming from the whole phase space in the calculation of the probability distribution.  
This is really worth analysing since in this case one would expect a competition between initial 
conditions coming from the regions of the available phase space where the system is ergodic and non-ergodic and 
therefore between Gaussian and $q$-Gaussian behaviour. Needless to say, as the region in the phase space where the 
system is non-ergodic diminishes (like the case $K=10$), Gaussian distribution will win, whereas the winner will 
be $q$-Gaussian as the region where the system is ergodic shrinks (like the case $K=0.2$). 
We notice that, if these regions coexist together, then this competition between Gaussian and 
$q$-Gaussian can be modelled as 

\begin{equation}
 \label{eq:compitition}
\frac{P(y)}{P(0)} =  \alpha\;  \exp_q(-\beta_q [yP(0)]^2) + (1-\alpha)  \exp(-\beta [yP(0)]^2)  .
\end{equation}
We check this hypothesis using our two appropriate cases, namely, $K=0.6$ and $K=2$. The results are 
given in Fig.~\ref{fig:k08-k2}, where a corroboration can be seen at different scales.

%%%%%%%%%%%%%%%%%%%%%%%%%%%%%%%
%%%%%%%%%%%%%%%%%%%%%%%%%%%%%%%%%%%%%

%%%%%%%%%%%%%%%%%%%%%%%%%%%%%%%%%%%
\section{Conclusions}
%%%%%%%%%%%%%%%%%%%%%%%%%%%%%%%%%%%
The phase space of the standard map presents regions of positive Lyapunov exponents 
coexisting with regions of zero Lyapunov exponents. The positive Lyapunov regions present 
mixing and thus the system is ergodic in those regions. 
For sufficiently low values of the control parameter $K$, the phase space is almost entirely 
dominated by zero Lyapunov behaviour and the distributions (obtained through time averaging, 
along the lines of central limit theorems) are $q$-Gaussians. 
As the value of $K$ increases the measure of the zero Lyapunov regions decreases, 
and we see a continuous crossing, expressed by the parameter $\alpha$ in Eq.~(\ref{eq:compitition}), 
between $q$-Gaussians distributions (Tsallis statistics) and Gaussian ones (Boltzmann-Gibbs statistics) 
with $\beta_q \to \infty$ and $\beta \to \pi$ as $K \to \infty$ (that is equivalent to $\alpha \to 0$). 
Remarkably enough, the distributions originated from initial conditions taken inside the region of islands, 
instead of over the entire phase space, yield one and the same value $q=1.935$, independently on 
whether we consider one or many of these regions, and independently from $K$. 
Initial conditions taken within the chaotic sea always yield Gaussians with $\beta = \pi$. 
The variance of $q$-Gaussians with $5/3<q<3$ diverges, though they have finite width. 
The $N$-fold convolution product of independent (or quasi-independent) $q$-Gaussians would 
asymptotically yield L\'evy distributions \cite{ct-levy}. Fig.~\ref{fig:lndistrib.k.02} neatly shows 
that this is {\it not} the case for the standard map: indeed, the actual time-averaging involves 
strong correlations.

%%%%%%%%%%%%%%%%%%%%%%%%%%%%%%%%
\section*{Acknowledgment}
%%%%%%%%%%%%%%%%%%%%%%%%%%%%%%%%
We thank I.\ L.\ S.\ Rodrigues for discussions during
the preliminary stages of this work. 
This work has been partially supported by TUBITAK (Turkish Agency) under the 
Research Project number 112T083. 
E.P.B. acknowledges
the National Institute of Science and Technology for Complex Systems,
and FAPESB through the program PRONEX (Brazilian agencies).
U.T. is a member of the Science Academy, Istanbul, Turkey.

%%%%%%%%%%%%%%%%%%%%%%%%%%%%%%%%%


\begin{thebibliography}{00}
%%%%%%%%%%%%%%%%%%%%%%%%%%%%%%%%%

\bibitem{CLT}
N. G. van Kampen, 
\textit{Stochastic Processes in Physics and Chemistry} 
(North Holland, Amsterdam, 1981). 

\bibitem{tsallis88}
C. Tsallis, 
J. Stat. Phys. {\bf 52}, 479 (1988). 

\bibitem{tsallisbook}
C. Tsallis, 
\textit{Introduction to Nonextensive Statistical Mechanics ---
Approaching a Complex World} (Springer, New York, 2009).


\bibitem{qCLT1}
S. Umarov, C. Tsallis, and S. Steinberg, 
Milan J. Math. {\bf 76}, 307 (2008).

\bibitem{qCLT2}
S. Umarov, C. Tsallis, M. Gell-Mann and S. Steinberg,
J. Math. Phys. {\bf 51}, 033502 (2010).



\bibitem{beck-lewis-swinney-2001}
C. Beck, G. S. Lewis, and H. L. Swinney,
Phys. Rev. E, {\bf 63}, 035303R (2001).

\bibitem{tsallis-borges-baldovin-2002}
C. Tsallis, E. P. Borges and F. Baldovin,
%``Mixing and equilibration: protagonists in the scene of 
%nonextensive statistical mechanics'',
Physica A {\bf 305}, 1 (2002).

\bibitem{renzoni1}
P. Douglas, S. Bergamini and F. Renzoni, 
Phys. Rev. Lett. {\bf 96}, 110601 (2006).

\bibitem{renzoni2}
E. Lutz and F. Renzoni, Nature Physics {\bf 9}, 615 (2013).

\bibitem{wilk}
C.-Y. Wong and G. Wilk, Phys. Rev. D {\bf 87}, 114007 (2013). 


\bibitem{betzler-borges-2012}
A. S. Betzler, and E. P. Borges,
Astronomy \&Astrophysics {\bf 539}, A158 (2012).

\bibitem{betzler-borges-2014}
A. S. Betzler, and E. P. Borges,
%\textit{Nonextensive Statistical Analysis of Meteor Showers 
%and Lunar Flashes},
Mon. Not. R. Astron. Soc. {\bf447}, 765 (2015).

\bibitem{cardone-leubner-delpopolo-2011}
V. F. Cardone, M. P. Leubner, and A Del Popolo,
Mon. Not. R. Astron. Soc., {\bf 414}, 2265 (2011).



\bibitem{tibet2005}
U. Tirnakli, C. Beck, and C. Tsallis, 
Physical Review E {\bf 75}, 040106R (2007).

\bibitem{tibet2009}
U. Tirnakli, C. Tsallis, and C. Beck,
Physical Review E {\bf 79}, 056209 (2009).

\bibitem{afsar1}
O. Afsar and U. Tirnakli,
EPL {\bf 101},  20003 (2013).


\bibitem{cirto-assis-tsallis-2014}
L. J. L. Cirto, V. R. V. Assis, and C. Tsallis,
Physica A {\bf 393}, 286 (2014).


\bibitem{Christodoulidi-Tsallis-Bountis-2014}
H. Christodoulidi, C. Tsallis, and T. Bountis,
Europhys. Lett., {\bf 108}, 40006 (2014).


\bibitem{anteneodo-tsallis-1998}
C. Anteneodo, and C. Tsallis, 
Phys. Rev. Lett. {\bf 80}, 5313 (1998).


\bibitem{ruiz}
G. Ruiz, T. Bountis and C. Tsallis,
Intern. J. Bifur. and Chaos {\bf 22}, 1250208 (2012).


\bibitem{ct-levy}
C. Tsallis, S. V. F. Levy, A. M. C. de Souza, and R. Maynard,
%``Statistical-mechanical foundation of the ubiquity of L\'evy distributions
%in nature'',
Phys. Rev. Lett., {\bf 75}, 3589 (1995); {\bf 77}, 5442 (erratum) (1996).
% pp.~3589--3593

\bibitem{ct-prato}
D. Prato, and C. Tsallis,
%``Nonextensive foundation of L\'evy distributions'',
Phys. Rev. E {\bf 60}, 2398 (1999). % pp. 2398--2401


\bibitem{chirikov}
B. V. Chirikov, Phys. Rep. {\bf 52}, 264 (1979).


\bibitem{zaslavsky1}
G. M. Zaslavsky, 
\textit{Hamiltonian Chaos and Fractional Dynamics} (Oxford University Press, 2005).


\bibitem{benettin}
G. Benettin, L. Galgani, A. Giorgilli, J. -M. Strelcyn, 
Meccanica {\bf 15}, 9 (1980).



\end{thebibliography}
\end{document}